\documentclass[twocolumn,prd,showpacs,amsmath]{revtex4}

\usepackage{graphicx}% Include figure files
\usepackage{dcolumn}% Align table columns on decimal point
\usepackage{bm}% bold math
\begin{document}

%\preprint{APS/123-QED}

\title{Upper limit on the ultra-high-energy photon flux from AGASA and
Yakutsk data}

\author{G.I.~Rubtsov$^1$, L.G.~Dedenko$^{2,3}$, G.F.~Fedorova$^3$,
E.Yu.~Fedunin$^3$, A.V.~Glushkov$^4$, D.S.~Gorbunov$^1$,
%M.~Kachelrie\ss$^{5,6}$,
I.T.~Makarov$^4$, M.I.~Pravdin$^4$,
T.M.~Roganova$^3$, I.E. Sleptsov$^4$
%K.~Shinozaki$^6$, M.~Teshima$^6$
and S.V.~Troitsky$^1$}

\affiliation{$^1$Institute  for Nuclear Research of the Russian
  Academy of Sciences,
  Moscow 117312, Russia}
\affiliation{$^2$Faculty of Physics, M.V.~Lomonosov Moscow State
  University,
  Moscow 119992, Russia}
\affiliation{$^3$D.V.~Skobeltsin Institute of Nuclear Physics,
  M.V.~Lomonosov Moscow State University,
  Moscow 119992, Russia}
\affiliation{$^4$Yu.G.~Shafer Institute of Cosmophysical Research and
  Aeronomy,
  Yakutsk 677980, Russia}
%\affiliation{$^5$Institutt for fysikk, NTNU Trondheim, N-7491
%  Trondheim, Norway}
%\affiliation{$^6$Max-Planck-Institut f\"ur Physik,
%  D-80805 M\"unchen,
%Germany}

%\date{muon2c.tex, last changed by mika on September 28, 2005}
%\date{muon2d.tex, last changed by kenjikry on Oct. 11, 2005}
\date{January 13, 2006}

\begin{abstract}
We present the interpretation of the muon and scintillation signals of
ultra-high-energy air showers observed by AGASA and Yakutsk extensive air
shower array experiments. We consider case-by-case ten highest
energy events with known muon content and conclude that at the
95\% confidence level (C.L.) none of them was induced by a primary
photon. Taking into account statistical fluctuations and differences in
the energy estimation of proton and photon primaries, we derive an upper
limit of 36\%  at 95\% C.L.\  on the fraction of primary photons in the
cosmic-ray flux above $10^{20}$~eV. This result disfavors the $Z$-burst
and superheavy dark-matter solutions to the GZK-cutoff problem.
\end{abstract}

\pacs{98.70.Sa, 96.40.De, 96.40.Pq}

\maketitle

%%%%%%%%%%%%%%%%%%%%%%%%% Introduction %%%%%%%%%%%%%%%%%%%%%%%%%%%%%%
\section{Introduction}
\label{sec:intro}
One of the most intriguing puzzles in astroparticle physics is
the observation of air showers initiated by particles with
energies beyond the cutoff predicted by Greisen and by Zatsepin and
Kuzmin~\cite{gzk}. Compared to lower energies, the energy losses of
protons
increase sharply at $\approx 5\times 10^{19}$~eV since pion production on
cosmic microwave background photons reduces the proton mean free path by
more than two orders of magnitude. This effect is even stronger for heavier
nuclei, while photons are absorbed
due to pair production on the radio background
with the mean free path of a few Mpc. Thus, the cosmic-ray (CR) energy
spectrum should dramatically steepen at $\approx 7\times 10^{19}$~eV for
any homogeneous distribution of CR sources.
Despite the contradictions in the shape of the spectrum,
the existence of air showers with energies
in excess of $10^{20}$~eV is firmly established by several independent
experiments using different
techniques (Volcano Ranch~\cite{exp}, Fly's
Eye~\cite{Bird}, Yakutsk~\cite{YakutskExperiment},
AGASA~\cite{agasares}, HiRes~\cite{HiRes} and Pierre Auger~\cite{Auger}
experiments). Some explanations for these showers, like the $Z$-burst or
top-down models, predict a significant fraction of photons above typically
$8\times 10^{19}$~eV (for reviews see, e.g., Refs.~\cite{reviews}).
Indications for the presence of neutral particles at lower energies were
found in Refs.~\cite{neutral}. Thus, the determination of the photon
fraction in the CR flux is of crucial importance, and the aim of this work
is to derive a stringent limit on this fraction in the integral CR flux
above $10^{20}$~eV. To this end, we compare the reported information on
signals measured by scintillation and by muon detectors for observed
showers with those expected by air shower simulations. We
focus on the surface detector signal density at 600 meters $S(600)$ (known
as charged particle density) and the muon density at 1000 m,
$\rho_{\mu}(1000)$, which are used in experiments as primary energy and
primary mass estimators, respectively.

We study individual events of
AGASA~\cite{AGASA_Eest} and of the Yakutsk extensive air shower array
(Yakutsk in what follows)~\cite{YakutskExperiment} with \textit{reconstructed}
energies above $8\times 10^{19}$~eV and measured muon content. We reject
the hypothesis that any of showers considered was initiated by a photon
primary at the 95\% confidence level (C.L.). We then derive as our main
result an upper limit of 36\% (at 95\%~C.L.)  on the fraction
$\epsilon_\gamma$ of primary photons with \textit{original} energies above
$10^{20}$~eV (the difference between original and reconstructed energies
is discussed in Sec.~\ref{sec:data}).

The rest of the paper is organized as follows. In Sec.~\ref{sec:data} we
discuss the experimental data set which we use for our study. In
Sec.~\ref{sec:simulations}, the details of the simulation of the artificial
shower libraries and comparison of the simulated and real data are given.
This section contains the description of our method and the main
results. We discuss how robust these results are with respect to
changes in assumptions, to analysis procedure, and to variations in the
experimental data, in Sec.~\ref{sec:robustness}. In
Sec.~\ref{sec:comparison}, we discuss the differences between our
approach and previous studies, which allowed us to put a
significantly more stringent limit on the gamma-ray fraction. Our
conclusions are briefly summarized in Sec.~\ref{sec:conclusions}.

%%%%%%%%%%%%%%%%%%%%%%%%% Experiments %%%%%%%%%%%%%%%%%%%%%%%%%%%%%%
\section{Experimental data}
\label{sec:data}
AGASA was operating from 1990 to 2003 and consisted of 111
surface scintillation detectors (covering an area of about $100$~km$^2$)
and 27 muon detectors. The areas of the AGASA muon detectors
varied between 2.8 and 20~m$^2$. The detectors consisted of 14--20
proportional counters aligned under a shield of either 30~cm of iron or
1~m of concrete and were placed below or close to scintillation detectors.
The threshold energy was 0.5~GeV$/\cos\theta_\mu$ for muons with zenith
angle $\theta_\mu$~\cite{AGASAmu}. During 14 years of operation, AGASA had
observed 11 events with reported energies above $10^{20}$~eV and zenith
angles $\theta<45^\circ$~\cite{agasares,AGASAarrDir}. Among them, six
events had $\rho_{\mu}(1000)$ determined~\cite{AGASAmu}.

Yakutsk is observing CRs of highest energies since 1973,
with detectors in various configurations. With
{$\theta< 60^\circ$}, it has observed three events
above $10^{20}$~eV, all with measured muon content. Before 1978, only
one muon detector with the area of 8~m$^2$ and threshold energy
$0.7$~GeV$/\cos\theta_\mu$ was in operation. Later, it has been
replaced by six detectors with areas up to 36~m$^2$ and the threshold
energy of $1.0$~GeV$/\cos\theta_\mu$~\cite{Yakutsk_mu}.

In our study, we combine the AGASA and Yakutsk datasets, motivated
by the following. First, both datasets are obtained from
surface array experiments operated with similar plastic scintillation
detectors. Second, the energy estimation procedures of the two
experiments are compatible, within the reported systematic errors at
$\sim 10^{20}$~eV, if differences in the observational conditions are
taken into account~\cite{SakakiThesis}. Finally, the values of the CR
flux at $10^{20}$~eV reported by the two experiments are consistent
within their $1\sigma$ errors.

%%%%%%%%%%%%%%%%%%%%%%%%% Methodology %%%%%%%%%%%%%%%%%%%%%%%%%%%%%%

The shower energy estimated by an experiment (hereafter denoted as
$E_{\rm est}$) is in general different
from the true primary energy (denoted as $E_0$) because of natural
shower fluctuations, etc.
Moreover, the energy estimation algorithms used by surface-array
experiments normally assume that the primary is a proton. While the
estimated energy  for nuclei depends only weakly on their mass number,
the difference between photons and hadrons is significant.
For photons, the effects of geomagnetic field~\cite{GMF} result in
directional dependence of the energy reconstruction. Thus,
the event energy reported by the experiment should be treated with
care when we allow the primary to be a photon. In this study we include
events with $E_{\rm est} \ge 8\times 10^{19}$~eV because of possible
energy underestimation for photon-induced showers; these events contribute
to the final limit, derived for $E_0>10^{20}$~eV, with different weights.

For AGASA, we use the events given in Ref.~\cite{agasares} that
pass the ``cut B'' defined in Ref.~\cite{AGASAmu}, that is having at
least one \footnote{We thank K.~Shinozaki for bringing a misprint in
Ref.~\cite{AGASAmu} to our attention: ``more than one'' was written
there.}
muon detector hit between 800~m and 1600~m from the shower axis. The
$\rho_\mu(1000)$ of the individual events can be read off from Fig.~2 of
Ref.~\cite{AGASAmu}. Yakutsk muon detectors have larger area and are more
sensitive both to weak signals far from the core and to strong signals for
which AGASA detectors might become saturated. This allowed the Yakutsk
collaboration to relax the cuts, as compared to AGASA, and to obtain
reliable values of $\rho_\mu(1000)$ using detectors between 400~m and
2000~m from the shower
axis~\cite{Knurenko,Yakutsk-muon-new}. Providing these cuts, six
AGASA and four Yakutsk events entered the dataset in our study (see
Table~\ref{events} for the event details).
\begin{table*}
\caption{\label{events}
Description of the individual events used in this work. Columns: (1),
event number; (2), experiment; (3), date of the event detection (in the
format dd.mm.yyyy); (4), the reported energy assuming a hadronic primary
(in units of $10^{20}$~eV); (5), the zenith angle (in degrees); (6) the
azimuth angle (in degrees, $\phi =0$ corresponds to a particle coming from
the South, $\phi =90^\circ$ -- from the West); (7) number of muon
detectors used to reconstruct muon density; (8) muon density at 1000~m from
the shower axis (in units of m$^{-2}$); (9), probability that this event
was initiated by a photon with $E>10^{20}$~eV; (10), probability that this
event was initiated by a non-photon with $E>10^{20}$~eV, assuming correct
energy determination. The sum $p_1^{(i)} +p_2^{(i)}$ gives the weight of
this event in the final limit on $\epsilon _\gamma $. The probability
that the primary had the energy $E<10^{20}$~eV is $1-p_1^{(i)}-p_2^{(i)}$.}
\begin{center}
\begin{ruledtabular}
\begin{tabular}{cccdddcddd}
$i$ & Experiment&Date & \multicolumn{1}{c}{$E_{\rm obs}$}
& \multicolumn{1}{c}{$\theta$} &
\multicolumn{1}{c}{$\phi$}&
\multicolumn{1}{c}{$n_{\rm det}$}&
\multicolumn{1}{c}{$\rho_{\mu}^{(i)}(1000)$}&
\multicolumn{1}{c}{$p_1^{(i)} $}& \multicolumn{1}{c}{$p_2^{(i)}$}\\
%\hline
(1)&
%\multicolumn{1}{c}{(1)}&
(2)&(3)&
\multicolumn{1}{c}{(4)}&
\multicolumn{1}{c}{(5)}&
\multicolumn{1}{c}{(6)}&
\multicolumn{1}{c}{(7)}&
\multicolumn{1}{c}{(8)}&
\multicolumn{1}{c}{(9)}&
\multicolumn{1}{c}{(10)}\\
\hline
1 &AGASA  & 10.05.2001 &2.46 & 36.5& 79.2 & 3&  8.9 &0.000 & 1.000\\
2 &AGASA  & 03.12.1993 &2.13 & 22.9& 55.5 & 1& 10.7 &0.001 & 0.998\\
3 &AGASA  & 11.01.1996 &1.44 & 14.2& 27.5 &$>1$&8.7 &0.013 & 0.921\\
4 &AGASA  & 06.07.1994 &1.34 & 35.1&234.9 & 1&  5.9 &0.003 & 0.887\\
5 &AGASA  & 22.10.1996 &1.05 & 33.7&291.6&$>1$&12.6 &0.000 & 0.581\\
6 &AGASA  & 22.09.1999 &1.04 & 35.6&100.0 &$>1$&9.3 &0.000 & 0.565\\
7 &Yakutsk& 18.02.2004 &1.60 & 47.7&180.8 & 5& 19.6 &0.000 & 0.876\\
8 &Yakutsk& 07.05.1989 &1.50 & 58.7&230.6 & 5& 11.8 &0.000 & 0.868\\
9 &Yakutsk& 21.12.1977 &1.10 & 46.1&346.8 & 1&  8.0 &0.000 & 0.645\\
10&Yakutsk& 02.05.1992 &0.85 & 55.7&163.0 & 5&  4.7 &0.000 & 0.303\\
\end{tabular}
\end{ruledtabular}
\end{center}
\end{table*}

\section{Simulations and results}
\label{sec:simulations}
\begin{figure}
\includegraphics[width=\columnwidth]{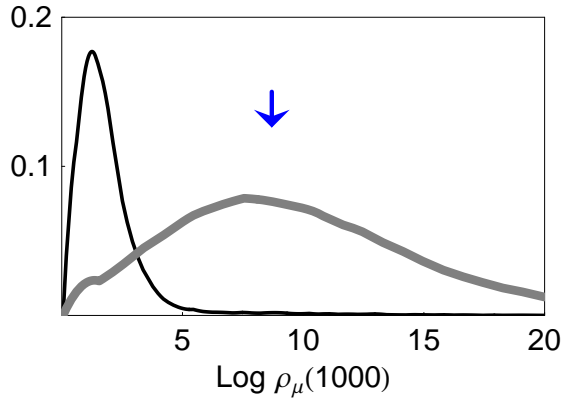}
\caption{\label{fig:event3}
Weighted distributions of muon density $\rho _\mu (1000)$ for the
simulated events compatible with the event 3 by $S(600)$ and the arrival
direction. Units in the vertical axis are arbitrary, $\rho _\mu (1000)$
is measured in m$^{-2}$. The thin dark line corresponds to primary photons;
it is the distribution used for our analysis. The thick grey line is the
distribution obtained in the same way but for 500 proton-induced showers.
The arrow indicates the observed value of $\rho _\mu (1000)$ for the event
3. The disributions include 50\% Gaussian error of the detector.}
\end{figure}

In order to interpret the data, for each of the ten events, we generated a
shower library containing 1000 showers induced by primary photons
\footnote{For the illustration in Fig.~\ref{fig:event3}, 500
proton-induced showers were simulated and processed in a similar way.}.
Thrown energies
%and 500 proton-induced showers
$E_0$ of the simulated showers were randomly selected (see below the
discussion of the initial spectra) between $5\times 10^{19}$~eV and $5\times
10^{20}$~eV to take into account possible deviations of $E_{\rm est}$ from
$E_0$. The arrival directions of the simulated showers were the same as
those of the corresponding real events. The simulations were performed
with CORSIKA~v6.204~\cite{Heck:1998vt}, choosing
QGSJET~01c~\cite{Kalmykov:1997te} as high-energy and
FLUKA~2003.1b~\cite{fluka} as low-energy hadronic interaction model.
Electromagnetic showering was implemented with EGS4~\cite{Nelson:1985ec}
incorporated into CORSIKA. Possible interactions of the primary photons
with the geomagnetic field were simulated with the PRESHOWER option of
CORSIKA~\cite{Homola:2003ru}. As discussed in
Sec.~\ref{sec:robustness:models}, this choice of the interaction models
results in a conservative limit on gamma-ray primaries. As suggested
in Ref.~\cite{Thin}, all simulations were performed with thinning level
$10^{-5}$, maximal weight $10^6$ for electrons and photons, and $10^4$ for
hadrons.

For each simulated shower, we determined $S(600)$ and $\rho_{\mu}(1000)$.
For the calculation of $S(600)$, we used the detector response functions
from Refs.~\cite{Sakaki,YakutskGEANT}.
For a given arrival direction, there is one-to-one correspondence
between $S(600)$ and the quantity called estimated energy,
$E_{\rm est}$. The relation is determined by
the standard analysis procedure of the two
experiments~\cite{AGASA_Eest,Yakutsk_Eest}.
This allows us to select simulated showers compatible with the
observed ones by the signal density. The quantity $S(600)$ is
reconstructed not precisely. In terms of estimated energy,
for AGASA events, the reconstructed energies are
are distributed with a Gaussian in $\log
\left(E_{\rm est}/\bar E_{\rm rec}\right)$;
the standard deviation of $E_{\rm est}$ is
$\sigma\approx 25\%$ \cite{SakakiThesis}. For Yakutsk events, the
corresponding $\sigma$ has been determined event-by-event and is typically
30--45\% \cite{PravdinICRC2005}.
To each simulated shower, we assigned a weight $w_1$ proportional to
this Gaussian probability distribution in $\log E_{\rm est}$ centered at
the observed energy $\bar E_{\rm rec}=E_{\rm obs}$.
Additionally, each simulated shower was weighted with $w_2$ to reproduce
the thrown energy spectrum $\propto E_0^{-2}$ which is typically predicted
by non-acceleration scenarios (see Sec.~\ref{sec:robustness:spectrum} for
a discussion of the variations of the spectral index). For each of the ten
observed events, we obtained a distribution of muon densities $\rho _\mu
(1000)$ representing photon-induced showers compatible with the observed
ones by $S(600)$ and arrival directions. To this end, we calculated $\rho
_\mu (1000)$ for each simulated shower by making use of the same muon
lateral distribution function as used in the analysis of real
data~\cite{AGASAmu,Yakutsk_mu}. To take into account possible experimental
errors in the determination of the muon density, we replaced each
simulated $\rho _\mu (1000)$ by a distribution representing possible
statistical errors (50\% and 25\% Gaussian for AGASA cut
B~\cite{AGASAmu50} and Yakutsk~\cite{Yakutsk-muon-new}, respectively). The
distribution of the simulated muon densities is the sum of these Gaussians
weighted by $w_1w_2$.

A typical distribution of simulated $\rho _\mu (1000)$ is given in
Fig.~\ref{fig:event3},
for gamma-
and proton-induced simulated showers compatible with the event 3 by
$S(600)$ and the arrival direction. We will see below that this particular
event has the largest probability of gamma interpretation among all ten
events in the data set; still the proton interpretation looks perfect for
it. This is the case for all events except event 7, which has too high
$\rho _\mu (1000)$ for a proton; possible nature of its primary particle will
be discussed elsewhere.

%%%%%%%%%%%%%%%%%%% Photon fraction limit %%%%%%%%%%%%%%%%%%%%%%%%%%%%%%

To estimate the allowed fraction $\epsilon_\gamma $ of primary
photons among CRs with $E_0>10^{20}$~eV,
we compare, for each observed event, two possibilities: (i)~that it was
initiated by a photon primary with $E_0>10^{20}$~eV and (ii)~that it was
initiated by any other primary with $E_0>10^{20}$~eV for which the
experimental energy estimation works properly.

Let us consider the $i$th observed event. Denote by $M$
the weighted number of showers contributed to the $\rho _\mu
(1000)$ distribution for the simulated photon-induced showers compatible
with the $i$th event by arrival direction and $S(600)$ (throughout
this paragraph, the weighted number is the sum of corresponding weights,
that is $M$ is the sum of weights of all 1000 showers simulated for the
$i$th event). Some of the simulated showers contributed to the part of the
distribution for which $\rho _\mu (1000)>\rho _\mu ^{(i)}(1000)$, where
$\rho _\mu ^{(i)}(1000)$ is the observed value for this event. The weighted
number of these showers is $M'$. Some part $l$ of this $M'$
corresponds to showers with
$E_0>10^{20}$~eV, the rest
($M'-l$) to $E_0<10^{20}$~eV. The probability
$p_1^{(i)}$ of case (i) is $p_1^{(i)}=l/M$, while the probability that the
event is consistent with a photon of $E_0<10^{20}$~eV is
$p_1^{\prime(i)}=(M'-l)/M$. Moreover, the probability that the event is
described by any other primary is
$1-p_1^{(i)}-p_1^{\prime(i)}=1-M^\prime/M$. We assume that the
experimental energy estimation works well for non-photon primaries and
determine the fraction $\xi$ of events with $E>10^{20}$~eV simply from the
Gaussian $\log(E_{\rm est})$ distribution, so the probability of the case
(ii) is $p_2^{(i)}=\xi(1-M^\prime/M)$. The values of $p_{1,2}^{(i)}$
are presented in Table~\ref{events}. Note that $p_1^{(i)}+p_2^{(i)}<1$
because of a non-zero probability that a simulated shower is initiated by
a primary with $E_0<10^{20}$~eV. This happens especially for events with
reported energies close to $10^{20}$~eV and reduces considerably the
effective number of events contributing to the limit on $\epsilon_\gamma$:
since we are interested in the limit for $E_0>10^{20}$~eV only, each event
contributes to the result with the weight $(p_1^{(i)}+p_2^{(i)})$.
Inspection of Table~\ref{events} demonstrates that the total effective
number of events with $E_0>10^{20}$~eV (the sum of $p_1^{(i)}$ and
$p_2^{(i)}$ over all ten events) is 7.67.

If the $i$th primary particle was a photon with $E_0>10^{20}$~eV with the
probability $p_1^{(i)}$ and a non-photon with $E_0>10^{20}$~eV with the
probability $p_2^{(i)}$, one can easily calculate the probability
${\cal P}(n_1,n_2)$ to have $n_1$ photons and $n_2$ non-photons in the set
of $N=10$ observed events ($0\le n_1+n_2 \le N$, the rest $N-n_1-n_2$
events have $E_0<10^{20}$~eV). From the set of $N$ events, one should take
all possible non-overlapping subsets of $n_1$ and $n_2$ events and sum up
probabilities of these realisations (since $p_{1,2}^{(i)}\ne
p_{1,2}^{(j)}$, these probabilities are different for different
realisations with the same $n_1$ and $n_2$).
Now, suppose that the fraction of the primary photons at $E_0>10^{20}$~eV
is $\epsilon _\gamma $. Then, the probability to have $n_1$ photons and
$n_2$ non-photons at  $E_0>10^{20}$~eV is $
\epsilon _\gamma ^{n_1} \left(1-\epsilon _\gamma
\right)^{n_2}
$,
and the probability that the observed muon densities were obtained with a
given $\epsilon _\gamma $ is
$$ {\cal P}(\epsilon _\gamma )=\sum\limits_{n_1,n_2=0}^N
 {\epsilon_\gamma} ^{n_1} \left(1-\epsilon_\gamma \right)^{n_2}
 {\cal P}(n_1,n_2) \,
$$
(cf.\ Ref.~\cite{Homola} for a particular case $n_1+n_2=N$; note that the
 combinatorial factor is included in the definition of ${\cal P}(n_1,n_2)$).
The cases $n_1+n_2<N$ reflect the possibility that some of
the $N$ events correspond to primaries with $E_0<10^{20}$~eV. In our
case, the probability ${\cal P}(\epsilon_\gamma)$ is a monotonically
decreasing function of $\epsilon_\gamma$. Thus the upper limit on
$\epsilon_\gamma$ at the confidence level $\alpha'$ is obtained by solving
the equation ${\cal P}(\epsilon_\gamma )=1-\alpha'$. For our dataset, the
95\%~C.L.\ upper limit on the photon fraction is $\epsilon_\gamma<0.33$.
The limit on $\epsilon_\gamma$ is rather weak compared to the
individual values of $p_1^{(i)}$ because of the small number of
observed events.

However, some of the photon-induced showers may escape from our
study because they may not pass the muon measurement quality cuts or
their estimated energy is below $8\times 10^{19}$~eV.
 Possible reasons for an underestimation of the energy may be either
the LPM effect~\cite{LPM} or substantial attenuation of gamma-induced
showers at large zenith angles.
To estimate the fraction of these
``lost'' events, we have simulated 1000 gamma-induced showers for each
experiment with arrival directions distributed according to the
experimental acceptance. We find that the fraction of the ``lost'' events
is $\sim 3.5\%$ for AGASA and  $\sim 15\%$ for Yakutsk. The account
of these fractions, weighted with the relative exposures of both
experiments, results in the final upper limit,
$$
\epsilon_\gamma <36\% ~~(95\%~{\rm C.L.}).
$$

\begin{figure}
\includegraphics[width=\columnwidth]{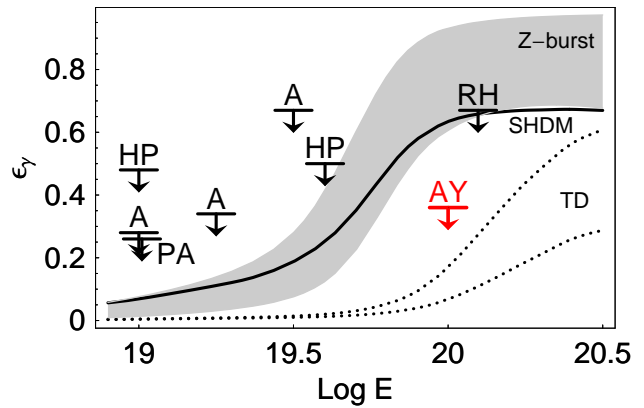}
\caption{\label{fig:limits}
  Limits (95\%~C.L.) on the fraction $\epsilon_\gamma$ of photons in the
  integral CR flux versus energy. The result of the present work
  (AY) is shown together with limits previously given in Refs.
  \cite{Haverah} (HP), \cite{AGASAmu} (A), \cite{Homola} (RH)
  and \cite{PAO} (PA). Also shown are predictions for the superheavy
  dark-matter (thick line) and topological-defect (necklaces, between
  dotted lines) models~\cite{ABK} and for the $Z$-burst model (shaded
  area)~\cite{SS}.}
\end{figure}

In Fig.~\ref{fig:limits}, we present our limit on $\epsilon_\gamma $
(AY) together with previously published limits on the same quantity.
Also, typical theoretical predictions are shown for the superheavy
dark-matter, topological-defect
and $Z$-burst models.
Our limit on $\epsilon_\gamma $ is  currently the
strongest one at $E_0>10^{20}$~eV. It disfavours some of the
theoretical models such as the $Z$-burst and superheavy dark-matter
scenarios.

\section{Robustness of the results}
\label{sec:robustness}
In this section, we discuss systematic uncertainties of our
limit that are related to the air shower simulations, to the data
interpretation and to selection cuts.
\subsection{Systematic uncertainty in the $S(600)$ and energy
determination}
\label{sec:robustness:energy}
The systematic uncertainty in the absolute energy
determination is 18\% and 30\% for AGASA~\cite{AGASA_Eest} and
Yakutsk~\cite{YakutskExperiment}, respectively. These systematic errors
originate from two quite different sources: (a)~the measurement of
$S(600)$ and (b)~the relation between $S(600)$ and primary energy.
The probabilities $p_1^{(i)}$ that a particular event may allow for a
gamma-ray interpretation are not at all sensitive to the
$S(600)$-to-energy conversion because we select simulated events by
$S(600)$ and not by energy. These probabilities may be affected by
relative systematics in determination of $\rho _\mu (1000)$ and $S(600)$.
On the other hand, in the calculation of $p_2^{(i)}$ we assumed that the
experimental energy determination is correct for non-photon primaries; the
values of $p_2^{(i)}$ and the effective number of events contributing to
the limit on $\epsilon _\gamma $ at $E_0>10^{20}$~eV would change if the
energies are systematically shifted. In our case (all $p_1^{(i)}\approx 0$),
the reported value of $\epsilon _\gamma $ would be applicable to the
shifted energy range in that case.

Thus, the 95\%~C.L. conclusion that none of the ten events considered here
was initiated by a photon is robust with respect to any changes in the
$S(600)$-to-energy conversion. As for the limit on $\epsilon _\gamma $ we
report, instead of $E_0>10^{20}$~eV, it would be applicable to a different
energy range if all experimental energies are systematically shifted. One
should note that theoretical predictions, e.g. the curves shown in
Fig.~\ref{fig:limits}, would also change because they are normalised to
the observed AGASA spectrum.

\subsection{Interaction models and simulation codes}
\label{sec:robustness:models}
Our simulations were performed entirely in the CORSIKA framework, and any
change in the interaction models or simulation codes, which affects either
$S(600)$ or $\rho _\mu (1000)$, may affect our limit. We have studied the
model dependence of our results by comparing different low- and high-energy
hadronic interaction models (GHEISHA~\cite{GHEISHA} versus FLUKA, SIBYLL
2.1~\cite{SIBYLL} versus QGSJET). Our
result is quite stable with respect to these changes. In all cases,
individual values of $p_1^{(i)}$
are always close to zero, thus the limit on $\epsilon_\gamma$ is
not affected. The change of the low energy model does not at all affect the
reported values. In use of SIBYLL compared with QGSJET, $\rho_\mu(1000)$
is $\sim 20\%$ smaller for photon-
induced showers. While $S(600)$ is almost unchanged,
events in our dataset are better explained by showers initiated by heavier
nuclei and the probability of photon-induced showers is even smaller. A
similar effect is expected for the coming interaction model
QGSJET~II~\cite{QGSJET-II}.

We also performed simulations with the help of
the hybrid code~\cite{hybrid} which reproduced the CORSIKA results with
high accuracy.
Another popular simulation code, AIRES~\cite{Aires}, differs from
CORSIKA mainly in the low-energy hadronic interaction model (which
is fixed in AIRES to be the Hillas splitting algorithm), hence we hope
that simulations with AIRES would not significantly affect our results.
Comparison with AIRES will be presented elsewhere.

The values presented here were
obtained for the standard parameterization of the photo-nuclear cross
section given by the Particle Data Group~\cite{PDG} (implemented as default
in CORSIKA). The muon content of gamma-induced showers is in principle
sensitive to the extrapolation of the photonuclear cross section to high
energies. The hybrid code~\cite{hybrid} allows for easy variations of the
cross section; we checked that the results are stable for various
reasonable extrapolations, in agreement with Ref.~\cite{RissePgamma}.
\subsection{Primary energy spectrum}
\label{sec:robustness:spectrum}
For our limit, we used the primary photon spectrum $E_0^{-\alpha }$ for
$\alpha =2$. While the individual probabilities $p_{1,2}^{(i)}$ are not
affected by the change of the spectral index $\alpha$ because the
simulated events are selected by $S(600)$ anyway, the value of $\alpha $
changes the fraction of ``lost'' photons and, correspondingly, the final
limit on $\epsilon _\gamma $. Variations of $1\le \alpha \le 3$ result in
the photon fraction limits between 36\% and 37\% (95\%~C.L.).
\subsection{Width of the $\rho _\mu $ distribution}
\label{sec:robustness:width}
Clearly, the rare probabilities of high values of $\rho _\mu (1000)$ in
the tail of the distribution for primary photons depend on the width of
this distribution. The following sources contribute to this width:
\begin{itemize}
 \item
variations of the primary energy compatible with the observed $S(600)$
(larger energy correspond to larger muon number and $\rho _\mu (1000)$);
\item
physical shower-to-shower fluctuations in muon density for a given energy
(dominated by fluctuations in the first few interactions, including
preshowering in the geomagnetic field);
 \item
artificial fluctuations in $S(600)$ and $\rho _\mu (1000)$ due to thinning;
 \item
experimental errors in $\rho _\mu (1000)$ determination.
\end{itemize}
While the first two sources are physical and are fully controlled by the
simulation code, the variations of the last two may affect the results.
\subsubsection{Artificial fluctuations due to thinning}
\label{sec:robustness:thinning}
It has been noted in Ref.~\cite{Badagnani} that
the fluctuations in $\rho _\mu (1000)$ due to thinning may affect strongly
the precision of the composition studies.
For the thinning parameters we use, the relative size of these fluctuations
is~\cite{our-thinning} $\sim 10\%$ for $\rho _\mu (1000)$ and $\sim
5\%$ for $S(600)$. Thus with more precise simulations, the distributions of
muon densities should become more narrow, which would reduce the
probability of the gamma-ray interpretation of each of the studied events even
further.
\subsubsection{Experimental errors in $\rho _\mu (1000)$
determination}
\label{sec:robustness:muon-errors}
The distributions of $\rho _\mu (1000)$ we use accounted for
the error in experimental determination of this quantity. The size of the
errors was taken from the original experimental
publications~\cite{Yakutsk-muon-new,AGASAmu50}. In principle, this error
depends on the event quality and on the muon number itself, which is lower
for simulated gamma-induced showers than for the observed ones. However,
e.g.\ Ref.~\cite{AGASAmu} states that for the AGASA cut A (two or more
muon detectors), the error is 40\%, lower than 50\% we
use~\cite{AGASAmu50}. Note that Ref.~\cite{AGASAmu} discusses muon
densities as low as 0.04~m$^{-2}$ and even 0~m$^{-2}$, much lower than
$\sim 1$~m$^{-2}$ typical for our simulated gamma-induced events. Still,
we tested the stability of our limit by taking artificially high values of
experimental errors in muon density: 100\% for AGASA and 50\% for
Yakutsk. The limit on $\epsilon _\gamma $ changes to 37\% (95\%~C.L.) in
that case.

\subsection{Data selection cuts}
\label{sec:robustness:cuts}
Since all events in the data set are unlikely to be initiated by
primary photons (all $p_1^{(i)}\approx 0$), the limit on $\epsilon _\gamma
$ is determined by statistics only and is affected if the number of events
is changed. Here, we discuss possible variations of the data set
corresponding to more stringent quality cuts which reduce the event number
and weaken the limit.
\subsubsection{Zenith angle}
\label{sec:robustness:ZA}
All Yakutsk events in the data set have zenith angles $45^\circ<\theta
<60^\circ$, so the cut $\theta<45^\circ$ imposed by AGASA reduces the
sample to six AGASA events which results in the limit $\epsilon _\gamma
<50\%$ (95\%~C.L.). One should note however that AGASA muon detectors are
not sensitive to inclined showers, which is not the case for Yakutsk.
\subsubsection{Core inside array}
\label{sec:robustness:event7}
Another cut imposed on the AGASA published dataset is the location of the
core inside array. The event number 7 does not satisfy this criterion; its
exclusion from the data set results in $\epsilon _\gamma <40\%$
(95\%~C.L.).
\subsubsection{More than one muon detector}
Reconstruction of the muon density at 1000~m from a single muon detector
reading requires extrapolation of the lateral distribution function with
an averaged slope. Though it is well-studied, the data points
corresponding to events with a single muon detector hit might be
considered less reliable than those with two or more hits. With the
account of the events with two or more hits only, we are left with seven events (four AGASA and
three Yakutsk) which weakens the 95\%~C.L.\ limit to $\epsilon _\gamma
<48\%$.
\section{Comparison with other studies}
\label{sec:comparison}
\begin{figure}
\includegraphics[width=\columnwidth]{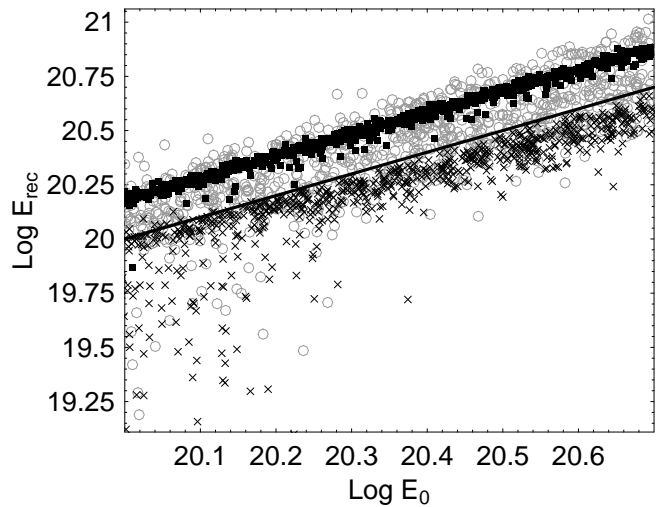}
\caption{\label{fig:2events}
Direction dependence of the reconstructed energy for gamma-ray primaries.
Plotted is the reconstructed energy (determined by the AGASA method from
$S(600)$) versus the primary energy. Dark boxes: arrival direction of
the event 1; crosses: arrival direction of the event 3; grey circles:
arrival directions randomly distributed according to the AGASA acceptance
($0<\theta<45^\circ$). Straight line represents $E_{\rm rec}=E_0$. Both
$E_0$ and $E_{\rm rec}$ are measured in eV.}
\end{figure}

Some of the previous studies used the AGASA~\cite{AGASAmu,Homola} and
Yakutsk~\cite{Knurenko} muon data to limit the gamma-ray primaries at high
energies. Our results differ from the previous ones not only because we
join the data sets of the two experiments. Two major distinctive
features of our approach allowed us to put the stringent limit:
\begin{itemize}
 \item
both $\rho _\mu (1000)$ and $S(600)$ were tracked for simulated showers
within framework of a {\em single } simulation code (CORSIKA in our case);
\item
each event was studied individually, without averaging over arrival
directions.
\end{itemize}
In Refs.~\cite{AGASAmu,Knurenko}, no conclusion was derived about
$\epsilon _\gamma $ at $E>10^{20}$~eV, and the data points corresponding
to highest-energy events were found to be quite close to the gamma-ray
domain. To our opinion, the main source of this effect is averaging over
arrival directions which introduced additional fluctuations for gamma-ray
primaries due to direction-dependent preshowering (see
Fig.~\ref{fig:2events} for an illustration).
In Ref.~\cite{Homola} which discussed the same six AGASA events,
all simulated showers for an event with the observed energy $E_{\rm obs}$
had energies
$1.2 E_{\rm obs}$ (up to the energy reconstruction uncertainty of 25\%).
This conversion had been obtained as the average over $\theta<36^\circ$ in
Ref.~\cite{AGASAmu} using AIRES simulation code~\cite{Aires}. That is, not
only the average results were applied to individual showers, but
effectively muon densities were simulated with CORSIKA while energies --
with AIRES, though the two codes result in a systematically different
relations between energy and $S(600)$. Artificially high energies resulted
in higher, closer to observed, muon densities for simulated photonic
showers. In our event-by event simulations with CORSIKA, the energies of
gamma-ray primaries whose $S(600)$ were compatible to observed values,
were not higher by a factor 1.2, but in fact even lower than $E_{\rm obs}$
for some of the events: besides the difference in simulation codes, this
is partially due to non-uniform distribution of the highest-energy AGASA
events on the celestial sphere~\cite{AGASAarrDir,NorthSouth} which makes
the usage of averaged energies poorly motivated.

The impact of two other sources of
difference between our approach and that of Ref.~\cite{Homola} is less
important for the final result: (i)~Ref.~\cite{Homola} does not account
for the ``lost'' photons and (ii)~the detector error is applied in our
study to the simulated events while in Ref.~\cite{Homola} -- to the
observed ones.

\begin{figure}
\includegraphics[width=\columnwidth]{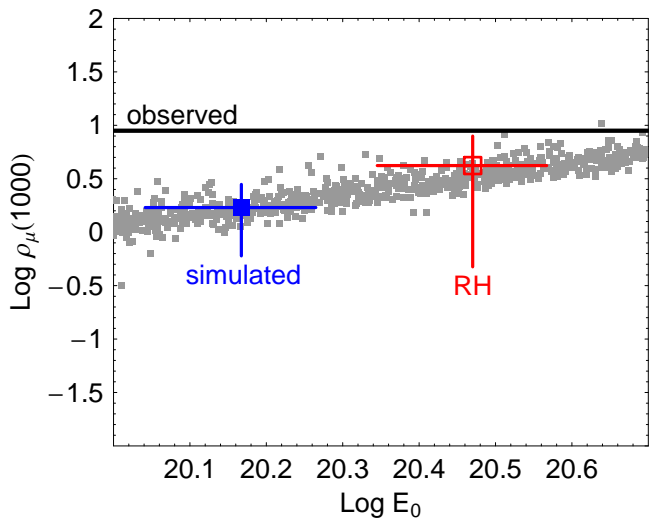}
\caption{\label{fig:DifferentEnergies}
Illustration of the difference between our study and Ref.~\cite{Homola}.
Plotted is the muon density at 1000~m versus the primary energy.
Small grey boxes: simulated gamma-induced events with arrival direction of
the event 1. Filled box, marked ``simulated'':
simulated events compatible with the event 1 by $S(600)$.
Open box, marked ``RH'': simulated
showers with average $E_0=1.2E_{\rm obs}$ from Ref.~\cite{Homola}.
The observed value of $\rho _\mu (1000)$ for the event 1 is
represented by a horizontal line, marked ``observed''.
$E_0$ is measured in eV, $\rho _\mu (1000)$ in m$^{-2}$. See the text
for more details.
}
\end{figure}

The difference with Ref.~\cite{Homola} is illustrated in
Fig.~\ref{fig:DifferentEnergies}, where $\rho _\mu (1000)$ is plotted
versus $E_0$ for simulated gamma-induced showers with the arrival
direction of the event \#1. For simulated events compatible with the real
event by $S(600)$, the average point is shown together with one sigma
error bars. Horizontal error bars correspond to variations in $E_0$
compatible with $S(600)$. Vertical error bars include variations in
simulated $\rho _\mu (1000)$ and 50\% detector error. The point
corresponding to simulated showers with $E_0=1.2E_{\rm obs}$ from
Ref.~\cite{Homola} has a larger $\rho _\mu (1000)$.
Horizontal error bars correspond to the energy reconstruction accuracy.
Vertical
error bars include variations in simulated $\rho _\mu (1000)$ reported in
Ref.~\cite{Homola} and 40\% detector error applied to the observed value,
added in quadrature.
We see that the main source of the
disagreement is in the values of $E_0$ which push, for the case of
Ref.~\cite{Homola}, the simulated muon densities closer to the observed
one.

\section{Conclusions}
\label{sec:conclusions}
To summarize, we have studied the possibility that the highest-energy
events observed by the AGASA and Yakutsk experiments were initiated by
primary photons. Comparing the observed and simulated muon content of
these showers, we reject this possibility for each of the ten events
at $E>8\cdot 10^{19}$~eV at least at the 95\% C.L.  An important
ingredient in our study is the careful tracking of differences between
the original and reconstructed energies.  This allows us to put an
upper bound of 36\% at 95\% C.L.\ on the fraction $\epsilon_\gamma $
of primary photons with original energies $E_0>10^{20}$~eV, assuming
an isotropic photon flux and $E_0^{-2}$ spectrum. This limit is the
strongest one up to date. It strongly disfavors the $Z$-burst and
constrains severely superheavy dark-matter models. The method that we
have used is quite general and may be applied at other energies and to
other observables.

We are indebted to
M.~Kachelrie\ss,
K.~Shinozaki and M.~Teshima
for numerous helpful discussions and collaboration at initial stages of
this work.
We thank L.~Bezroukov, V.~Bugaev, R.~Engel, D.~Heck, A.~Ringwald, M.~Risse,
V.~Rubakov, D.~Semikoz and P.~Tinyakov for helpful discussions and
comments on the manuscript. This study was
performed within the INTAS project 03-51-5112. We acknowledge also support
by fellowships of the Russian Science Support Foundation and of the
Dynasty foundation (D.G.\ and S.T.), by the grants NS-2184.2003.2 (D.G.,
G.R.\ and S.T.), NS-1782.2003.2, RFFI 03-02-16290 (L.D., G.F., E.F.\ and
T.R.), NS 748.2003.2, RFFI 03-02-17160, RFFI 05-02-17857, FASI
02.452.12.7045 (A.G., I.M., M.P.\ and I.S.). G.R.\ and S.T.\ thank the
Max-Plank-Institut f\"ur Physik (Munchen), where a significant part of
this work was done, for warm hospitality. Computing facilities of the
Department of Theoretical Physics, Institute for Nuclear Research
(Moscow), were used to perform the simulations of air showers.

%%%%%%%%%%%%%%%%%%%%%%%%%%%%%%%%%%%%%%%%%%%%%%%%%%%%%%%%%%%%%%%%%%

\end{document}